# Design of Information and Telecommunication Systems with the Usage of the Multi-Layer Graph Model

Dmytro Ageyev[1], Artem Ignatenko[2], Fouad Wehbe[3]

1. TCS Department, Kharkiv National University of Radioelectronics, UKRAINE, Kharkiv, Lenina ave. 14, E-mail: dm@ageyev.in.net

2. TCS Department, Kharkiv National University of Radioelectronics, UKRAINE, Kharkiv, Lenina ave. 14, E-mail: sanitarium@ukr.net

3. TCS Department, Kharkiv National University of Radioelectronics, UKRAINE, Kharkiv, Lenina ave. 14, E-mail: fouadera@mail.ru

*Abstract* – **Considered in the paper is the overlaid nature of modern telecommunication networks and expediency of usage the multi-layer graph model for its design. General statement of design problem is given. The design method in general using multi-layer graph model and its application for a typical information and telecommunication system is proposed.**

## I. Introduction

Modern telecommunication systems are constructed on the basis of overlay networks (i.e. each lower level of the network provides transparent transfer of the network flow over the upper level); thus, they have a multi-layer structure formed with topology hierarchy. When solving design problems, it is necessary to determine the network structure on each layer.

An important objective in modern information and communication systems design is to provide a quality of service. In [1], the authors solved the problem of the distribution of flows using a tensor model for a multipath QoS-routing.

Known approaches for solving design problems use the step-by-step synthesis separately for each level to consider the multi level nature of systems. Synthesis results of one level are the source data for the rest of the levels. In this case the interconnections and interdependences between the levels are not considered. As the result, the final configuration is not optimal.

To solve this problem, the authors propose [2, 3] to use the multi-layer network model represented as an ordered set of graphs. Topology of each graph can be different; they may have different sets of edges; in this case, as a rule, the set of upper layer nodes is a subset of the lower layer nodes.

The above model of the multi-layer network structure has an accurate correspondence of nodes in graphs that describe each layer. At the same time, a lot of systems have interlevel connections of a more complex nature, for which the above model is no longer adequate. To eliminate this drawback, the model based on multi-layer graph is proposed [4, 5, 6].

## II. Problem Description

Generally, we can describe the statement of information and telecommunication network design problem as follows:

$U = \{u_i\}$ - set of subscribers;

$S = \{s_i\}$ - set of servers;

$Z = \{z_i\}$ - set of intermediate nodes;

$B = \{b_i\}$ - set of channels;

$C(b_i)$ - bandwidths of channels;

$P(s_i)$ - productivity of servers;

As the result of structural and parametric synthesis, it is necessary to define a subset of subscribers assigned to each of the servers $U_A(s_i) \subset U$, a subset of channels included in the resulting network topology $B_A \subset B$, flow transfer routes in order to provide the minimum of summary amount of the flows being transferred in the network:

$$\sum_i \gamma(b_i^A) \to \min.$$

## III. Modeling of information and telecommunication system with multi-layer graph

Graphs are widely used for the mathematical modeling of the structure of systems including telecommunication systems. When the system is described using a graph, the system elements are modeled as nodes, and connections between them are modeled as arcs or edges.

For overlay networks modeling, we propose to use multi-layer graph $MLG = (\Gamma, V, E)$, which includes:

- a set of subgraphs $\Gamma = \{\Gamma^1, \ldots \Gamma^l, \ldots, \Gamma^L\}$, where the subgraph $\Gamma^l = (V^l, E^l)$ describes the network structure on the layer $l$;

- nodes $v_i \in V$ and edges $e_k = (v_i, v_j)$, $e_k \in E$ provide the interconnection between subgraphs $\Gamma^l$.

The structure of graph MLG modeling telecommunication networks is subject to an additional constraint which reads that for each edge $e_k^l = (v_i^l, v_j^l)$, $e_k^l \in E^l$ of subgraph $\Gamma^l$ there exists a path $\pi = (v_i^l, \ldots, v_m^n, \ldots v_j^l)$ between the nodes $v_i^l$ and $v_i^l$, $v_i^l, v_j^l \in V^l$, that passes through the lower layer graph:

$$\forall e_k^l = \left(v_i^l, v_j^l\right), \quad e_k^l \in E^l, \quad v_i^l, v_j^l \in V^l,$$
$$\exists \pi = \left(v_i^l, \ldots, v_m^n, \ldots, v_j^l\right), v_m^n \in V^n, n < l.$$

This rule does not apply only for the lowest layer subgraph, $l = 1$.

Describing telecommunication networks with the multi-layer graph allows considering technological hierarchy of modern networks, specifically, overlaid principle of their construction in contrast to classic graphs.

The description of telecommunication system with multi-layer graph is done in accordance with the following strategy.

Step 1. Distinguish the set of layers in the modeled telecommunication system.

Step 2. Describe the topology of each layer with a classic graph.

Step 3. Distinguish logical, functional and physical connections between the objects on the different layers and describe them with graphs.

Step 4. Assign the edges and nodes a set of parameters, describing the parameters of the respective objects and connections that are of interest for modeling.

On step 1 we distinguish a set of overlay structures in the modeled system. They can act as a separate object for analysis.

On step 2 we define the network topology for each separate layer distinguished on step 1. For this:
- every node of the network on this level is represented as a graph node.
- pairs of directly interconnected nodes are determined (i.e. pairs of nodes that do not use other nodes of this layer as transit nodes to interconnect between each other).
- for each pair of the directly interacting nodes, we include an edge into the graph that connects the corresponding graph vertices.

On step 3, each edge included into the multi-layer graph corresponds to the logical and physical connection between the nodes of the overlay networks.

On step 4, each edge of the multi-layer graph is ascribed a number of parameters corresponding to the parameters of the network being modeled.

### III. SOLUTION APPROACH

General methods of the informational telecommunications systems design consist of the following steps:

Step 1. Analysis of the subject problem and synthesis of the initial redundant multilayer graph describing the topological and functional structure of the system being designed.

Step 2. Statement of the optimization problem of the telecommunication system design.

Step 3. Solving of the optimization problem (search for the optimal multi-layer sub-graph for the initial redundant multilayer graph).

Step 4. Interpretation of the multi-layer graph achieved on the previous step as a project solution.

Let us consider these steps in more detail.

Actions performed on step 1 are based on the description methods of structural and functional systems properties with the usage of a multilayer graph.

On step 3, we seek the optimal multi-layer graph, and if need be, the values of parameters associated with the vertices and edges. On this stage, depending on the mathematical problem statement, we apply the methods of combinatory optimization, and the methods of linear and non-linear programming.

On step 4, we apply the procedures which are contrary to those in step 1, and they result in the fact that the achieved structure of the multilayer graph is transformed into the project solution interpretable in terms of the object domain (specialization of the exploited equipment, parameters of its components, protocols used).

Using the general methods mentioned above, the design of a typical informational telecommunication system can be described in the following way [7].

1. In order to describe the system with a multilayer graph G, let us distinguish its three levels (subgraphs $\Gamma^1$, $\Gamma^2$, $\Gamma^3$):
- the level of cooperation between the service and subscribers ($\Gamma^3$);
- the level of cooperation between the subscribers and the servers, and the servers with each other ($\Gamma^2$);
- the level of physical topology of the system being modeled, the level of physical interaction between the subscribers, access nodes and servers ($\Gamma^1$).

2. On the level of interaction between the subscribers and the service, the vertices $a_i^3$ of graph $\Gamma^3$ correspond to the network subscribers (or groups of subscribers connected to the same access node). The vertex $v_0^3$ models the service being provided. If the service is provided to all groups of subscribers within the system, the graph $\Gamma^3$ will have the "star" topology:

$$e_i^3 = \left(v_0^3, a_i^3\right), \quad e_i^3 \in E^3, \quad i = 1, \ldots, N_A,$$

where $N_A$ is the number of the network subscribers (groups of subscribers).

On the level of interaction between the subscribers and the servers, and the servers with each other, the topology will approach to mesh topology because it is initially envisaged that each server can interact with the other servers, and each subscriber can receive content from any server. Besides this, the subscribers have no connection with each other, and consequently, there will be no edges between the vertices corresponding to the subscribers.

$$\forall v_i^2, a_j^2, \quad \exists e^2 = (v_i^2, a_j^2),$$
$$\forall v_i^2, v_j^2, \quad i \neq j \quad \exists e^2 = (v_i^2, v_j^2),$$
$$\forall a_i^2, a_j^2, \quad !\exists e^2 = (a_i^2, a_j^2)$$

Finally, the lowest level describes the physical interaction between the servers, subscribers and access nodes.

3. The multi-level graph concept implies the modeling of overlay networks – correspondingly, every functional unit on each level has a reflection in the adjoining level. Such reflection on a multi-level graph is expressed by the set of edges $E = \{e_i\}$, connecting the functional units (nodes) that correspond to each other on the adjoining levels of the multi-level graph.

4. In order to describe the functional characteristics of the system being modeled, let us use the flow model on the multi-level graph [8]. On this stage, for simplification purposes, we will only operate the value of the amount of flow disregarding the delays and other parameters of the telecommunication systems elements. A flow being transferred over the edge connecting a group of subscribers with a service equals the sum of flows of all the sessions initialized by the subscribers:

$$\gamma(e_i^3) = \sum_n \gamma(S_n(a_i^3))$$

When the vertex $v_0^3$ is projected upon the set of servers $V^2$ the following condition should be observed:

$$\sum_i P(v_i^2) = P(v_0^3),$$

where $P(N)$ is the node productivity.

At this point, the bandwidths of the edges connecting the vertices of the graphs $\Gamma^2$ and $\Gamma^3$ corresponding to the network subscribers are not limited, while for the edges connecting the vertex $v_0^3, v_0^3 \in \Gamma^3$ and vertices $v_i^2, v_i^3 \in \Gamma^3$ the bandwidth equals the productivity of the server, upon which the following is projected:

$$c(e(a_i^3, a_i^2)) = \infty,$$
$$c(e(v_3^0, v_i^2)) = P(v_i^2)$$

Bandwidths of the edges $e_{ij}^1 = (v_i^1, v_j^1)$, $e_{ij}^1 \in \Gamma^1$ on the lowest layer of the graph $\Gamma^1$ should be more or equal to the amount of flow being transferred over them:

$$c(e_{ij}^1) \geq \gamma(e_{ij}^1).$$

Now it is necessary to describe the flow model on the multi-layer graph [8]. The process of information flows transfer through the network is modeled by the flow transfer over the graph edges. If we use a multi-layer graph, the edges can be divided into two groups:
- edges connecting the nodes of the same layer;
- edges connecting the nodes on different layers of a multi-layer graph.

In the first case, the edges model the processes occurring on the same layer of the network. In the second case, they model the influence of the processes occurring on one level, upon the processes occurring on another level.

At this point, the flow conservation law should be observed.

It can be briefly formulated in the following statements:
- amount of flow being transferred between a pair of interacting nodes (source – destination) over the chosen path is always the same;
- sum of flows being transferred over different paths between a pair of interacting nodes (source – destination) equals the amount of requirements occurring in the source node and equals the amount of requirements being processed in the destination node;
- sum of flows incoming to the node equals the sum of flows outgoing from the node if the node functions only as a transit node; or it differs by the amount equaling to the difference between the amounts of flows for which it is a source or a destination.

The application of the specified flow model allows to consider the connections between the flow amounts on different levels and the limitation of the system elements bandwidth (productivity).

The lowest level subgraph $\Gamma 1$ is a redundant graph including all intermediate nodes and access points. Its redundancy enables us to solve the following problems:
- network topology synthesis on the redundant graph;
- finding of a path with the given characteristics on a graph of existing topology.

The network design problem can be described in two statements depending on the source data and sought-for parameters.

In the first case, it is necessary do define, for the given amount of queries, what resource should be allocated and how it should be cost-efficiently distributed in the network considering the constraining conditions for the routing and the flow. This problem is known (for example, in [9]) as uncapacitated design. As a rule, this problem is actual in the middle-term and long-term perspective of the network design.

However, as soon as the limited resulting network resource and the amount of queries are known, the problem transforms itself into the problem of flow distribution over different paths so that the solution should be optimal in accordance with the given criteria (e.g. minimum cost or maximum profit). This statement is known as capacitated problem, which is actual in the short-term perspective of the network design when the network resource cannot be increased.

The majority of similar problems can be formulated as the problem of transfer of multi-product flow on the basis of paths (link-path formulation), or on the basis of the nodes (node-link formulation).

Applying the linear or linear integer programming, we can achieve a resulting graph consisting only of the edges that will carry the traffic, i.e. the set of channels $B = \{B_{ij}\}$.

The topology synthesis problem on the redundant graph, in this case, is reduced to the flows distribution on a multi-layer graph in a way so that a minimum weight graph could be found taking into account the restrictions to the edges bandwidth.

## IV. Conclusions

Modern information telecommunication systems have a multilayer structure, which can be divided into two types: organizational structure and technological one. Organizational structure is characteristic when territorial and distribution fragments of the network performing different functions (WAN, MAN, LAN) are distinguished. Technological multilevel structure is composed by overlay networks.

The design of modern multiservice telecommunication systems requires the definition of its structure simultaneously on its several levels formed by overlay networks. In order to solve this problem as a complex, we propose for the first time to use the mathematical model of multi-layer graph.

Consequent solving of the design problem for each of the network levels separately, when the results of the design on one level become the source data for the remaining network levels, does not consider the system as a whole, but only provides an optimal result for each sublevel, which can lead to falling in the local optimums but not optimally solve the problem as a whole. This drawback can be eliminated by applying the mathematical model of multilayer graph.

Mathematically, the multilayer graph is an object of theory of sets allowing to reflect the key properties of the overlay networks. The multilayer graph is a further development of classical graphs, but contrary to them consists of multiple graphs called layers, and of a graph connecting these layers. Besides this, for each graph edge of a upper level there should exist a path in the graph of a lower level.

The flow model, which takes into account the flows distribution over the traffic transfer routes and allows to define their values for different system elements, is widely used for the problem statement of telecommunication systems design. The flow model is also widely used for the formulation of coherence conditions of the network being modeled.

## References

[1] O.V. Lemeshko and O.A. Drobot, "Mathematical Model of Multipath QoS-based Routing in Multiservice Networks", in *VIII$^{th}$ International Conference on Modern Problems of Radio Engineering, Telecommunications, and Computer Science, TCSET'2006*. Lviv Polytech. Nat. Univ., Lviv-Slavske, Ukraine, 2006, pp. 72-74.

[2] S. Orlowski, A.M.C.A. Koster, C. Raack and R. Wessäly "Two-layer network design by branch-and-cut featuring MIP-based heuristics" in *3$^{rd}$ International Network Optimization Conference (INOC 2007)*, Spa, Belgium, 2007, pp. 114–119.

[3] A. Capone, G. Carello and R. Matera, "Multi-Layer Network Design with Multicast Traffic and Statistical Multiplexing", in *IEEE Global Telecommunications Conference (IEEE GLOBECOM)*, Washington, USA, 2007, pp. 2565–2570.

[4] D.V. Ageyev "Metodika opisaniya struktury sovremennykh telekommunikatsionnykh sistem s ispol′zovaniem mnogosloinykh grafov" [The method of describing the structure of modern telecommunication systems using multi-layer graph], *Eastern European Journal of Enterprise Technolopgies*, No 6/4 (48), pp. 56-59, 2010.

[5] D.V. Ageyev, "Modelirovanie sovremennykh telekommunikatsionnykh sistem mnogosloinymi grafami" [Simulation of modern telecommunication systems with multi-layer graphs usage], *Problemi telekomunìkacìj*, No 1(1), pp. 23–34, 2010, http://pt.journal.kh.ua/2010/1/1/101_ageyev_simulation.htm

[6] Haidara Abdalla and D. V. Ageyev, "Application of Multi-layer Graphs In the Design of MPLS Networks," in *XI$^{th}$ International Conference on Modern Problems of Radio Engineering, Telecommunications, and Computer Science, TCSET'2012*, Lviv Polytech. Nat. Univ., Lviv-Slavske, Ukraine, 2012, pp. 336–337.

[7] D. V. Ageyev and A. A. Ignatenko, "Describing and Modeling of Video-on-Demand Service with the Usage of Multi-Layer Graph," in *XI$^{th}$ International Conference on Modern Problems of Radio Engineering, Telecommunications, and Computer Science, TCSET'2012*, Lviv Polytech. Nat. Univ., Lviv-Slavske, Ukraine, 2012, pp. 340 – 341.

[8] D.V. Ageyev, "Metod proektirovaniya telekommunikatsionnykh sistem s ispol′zovaniem potokovoi modeli dlya mnogosloinogo grafa" [Method of designing telecommunication systems using flow model for multi-layer graph], *Problemi telekomunìkacìj*, No2(2), pp. 7 – 22, 2010, http://pt.journal.kh.ua/2010/2/2/102_ageyev_layer.htm.

[9] Micha Pióro and Deepankar Medhi *Routing, Flow, and Capacity Design in Communication and Computer Networks*. Elsevier, 2004